# Spatial and polarization division multiplexing harnessing on-chip optical beam forming

David González-Andrade,[1,*] Xavier Le Roux,[1] Guy Aubin,[1] Farah Amar,[1] Thi Hao Nhi Nguyen,[1] Paula Nuño Ruano,[1] Thi Thuy Duong Dinh,[1] Dorian Oser,[1,2] Diego Pérez-Galacho,[3] Eric Cassan,[1] Delphine Marris-Morini,[1] Laurent Vivien,[1] and Carlos Alonso-Ramos[1]

[1]*Centre de Nanosciences et de Nanotechnologies, CNRS, Université Paris-Saclay, Palaiseau 91120, France*
[2]*QuTech and Kavli Institute of Nanoscience, Delft University of Technology, Delft 2600 GA, The Netherland*
[3]*Telecommunication Research Institute (TELMA), Universidad de Málaga, CEI Andalucía TECH, Málaga 29010, Spain*
**Corresponding author: david.gonzalez-andrade@c2n.upsaclay.fr*

**On-chip spatial and polarization multiplexing have emerged as a powerful strategy to boost the bandwidth of integrated optical transceivers. State-of-the-art multiplexers require accurate control of the relative phase or the spatial distribution among different guided optical modes, seriously compromising the bandwidth and performance of the devices. To overcome this limitation, we propose a new approach based on the coupling between guided modes in integrated waveguides and optical beams free-propagating on the chip plane. The engineering of the evanescent coupling between the guided modes and free-propagating beams allows spatial and polarization multiplexing with state-of-the-art performance. To demonstrate the potential and versatility of this approach, we have developed a two-polarization multiplexed link and a three-mode multiplexed link using standard 220-nm-thick silicon-on-insulator technology. The two-polarization link shows a measured -35 dB crosstalk bandwidth of 180 nm, while the three-mode link exhibits a -20 dB crosstalk bandwidth of 195 nm. These bandwidths cover the S, C, L, and U communication bands. We used these links to demonstrate error-free transmission (bit-error-rate < $10^{-9}$) of two and three non-return-to-zero signals at 40 Gbps each, with power penalties below 0.08 dB and 1.5 dB for the two-polarization and three-mode links respectively. The approach demonstrated here for two polarizations and three modes is also applicable to future implementation of more complex multiplexing schemes.**

## 1. INTRODUCTION

Silicon photonics has been identified as a promising technology to address the communication bottleneck in data centers and long-haul networks [1]. On-chip optical transceivers fabricated at large volume using microelectronics facilities could become instrumental in exploiting optical carriers to boost the data bandwidth while reducing the power consumption of communication systems [2,3]. Current silicon photonics optical transceivers carry different data channels at distinct wavelengths using wavelength-division multiplexing (WDM) [4]. However, as the industry moves forward with the development of next-generation optical networks, other multiplexing schemes will be required to support a higher data capacity per wavelength channel. One promising approach is the use of orthogonal polarizations and spatially-distributed modes to encode more data channels at a specific wavelength [5]. The manipulation of higher-order modes and polarization state of light on chip has thus attracted a significant research interest in the past years for the realization of performant mode and polarization (de)multiplexers.

A myriad of architectures has been proposed to realize mode-division and polarization-division multiplexing. Mode-division multiplexing (MDM) has been demonstrated based on multimode interference (MMI) couplers, asymmetric Y-junctions and directional couplers (ADC), adiabatic tapers, pixelated-meta structures, and subwavelength metamaterials [6-13]. Similarly, multiple devices have been reported for polarization-division multiplexing (PDM), including MMIs, Mach-Zehnder interferometers, diverse types of directional couplers (i.e., symmetric, asymmetric, tapered and bent), photonic crystals, slot waveguides, and subwavelength and tilted nano-gratings [14-24]. Despite the great diversity of proposed solutions, crosstalk still remains as one of the main impairments in high-speed optical communications, especially in systems with a high-channel count. An increased crosstalk has a negative impact on the link performance in terms of bit-error-rate (BER) and ultimately results in a power penalty that jeopardizes low power consumption [25]. Silicon multiplexers handling more than two modes (e.g., 3 spatial modes with the same polarization) yield poor crosstalk values ranging from -19 dB [11] to -9.7 dB [7].

Here, we propose a simple yet effective strategy to realize highly efficient mode and polarization (de)multiplexing based on engineered on-chip beam forming. Instead of controlling the phase or field distribution matching between two guided modes, we engineer the evanescent coupling between the modes of a photonic waveguide and free-propagating beams on the chip plane. Coupling between guided modes and on-chip free-propagating beams has been achieved using distributed Bragg deflectors [26-30]. However, the wavelength-dependent nature of the diffractive coupling seriously limits their use for wideband mode or polarization multiplexers. Conversely, evanescent coupling does not present a strong wavelength dependence, which allowed the demonstration of on-chip beam expanders with a wide bandwidth [31-33]. In our proposed scheme we engineer the evanescent coupling to make each waveguide mode (or polarization) couple to a different in-plane beam, propagating with a specific angle. As schematically shown in Fig. 1(a), this strategy spatially separates the different waveguide modes, allowing (de)multiplexing. Based on this approach, we experimentally demonstrate a two-polarization link and a three-mode link that allow error-free transmission of multiplexed high-speed data streams. We show transmission of two and three 40 Gbps non-return-to-zero (NRZ) signals, with negligible power penalties at a BER of $10^{-9}$. To the best of our knowledge, this is the first demonstration of mode and polarization handling enabled by beamforming in an integrated circuit. These devices could be a promising alternative to fixed layout-guided architectures, with excellent potential for the next generation of high-speed and large-capacity on-chip optical interconnects.

## 2. MODE DIVISION MULTIPLEXING

### A. Operation principle

Figure 1(a) shows the top view of a waveguide evanescently coupled to an adjacent slab. The input waveguide has a width of $W$ and is placed at a distance $G$ from the slab. The waveguide supports a discrete set of guided modes with effective indices of $n_{eff}^{m}$, with $m$ being a natural number indicating the mode order. The slab supports a continuum of vertically-confined ($z$-axis) modes that propagate freely in the $xy$-plane with a wavenumber given by:

$$\overrightarrow{\boldsymbol{k_s}} = \boldsymbol{k_0 n_s}[\mathbf{sin}(\boldsymbol{\theta})\hat{\boldsymbol{x}} + \mathbf{cos}(\boldsymbol{\theta})\hat{\boldsymbol{y}}], \quad (1)$$

where $n_s$ is the effective index of the slab, $\theta$ is the propagation direction angle of the slab mode with respect to the $y$-axis, $k_0 = 2\pi/\lambda_0$ is the vacuum wavenumber, and $\hat{x}$ and $\hat{y}$ are the unitary vectors in the $x$- and $y$-directions, respectively. Phase-matching between the guided waveguide modes and the slab modes occurs for

$$\mathbf{sin}(\boldsymbol{\theta_m}) = \frac{\boldsymbol{n_{eff}^{m}}}{\boldsymbol{n_s}}. \quad (2)$$

Note that Eq. (2) is analogous to the grating equation when the zero-order harmonic is considered [34,35]. This zero-order operation obviates the strong wavelength dependence of the propagation angle in distributed Bragg deflectors [26-30]. The effective indices of the waveguide ($n_{eff}^{m}$) and slab ($n_s$) are wavelength dependent. However, both values are inversely proportional to the wavelength, reducing the wavelength dependence of the ratio $n_{eff}^{m}/n_s$. If the separation distance $G$ is sufficiently small, the waveguide mode with effective index $n_{eff}^{m}$ will evanescently couple to the slab mode propagating with an angle $\theta_m$, satisfying Eq. (2). After a given propagation distance, the different slab modes are spatially separated allowing demultiplexing. Due to reciprocity, the same device can be used to multiplex different slab beams into different waveguide modes.

For the implementation of the proposed architecture, we consider silicon-on-insulator (SOI) with a thickness of the silicon guiding layer of 220 nm. Figure 1(b) shows the effective indices of the $TE_0$, $TE_1$, and $TE_2$ modes calculated as a function of the waveguide width $W$. Effective indices are calculated using a commercial finite difference eigenmode (FDE) solver. Figure 1(c) shows the propagation angle of the slab modes satisfying Eq. (2), considering a slab index $n_s = 2.85$. Differences in propagation angle exceeding 15° can easily be achieved by properly choosing the waveguide width.

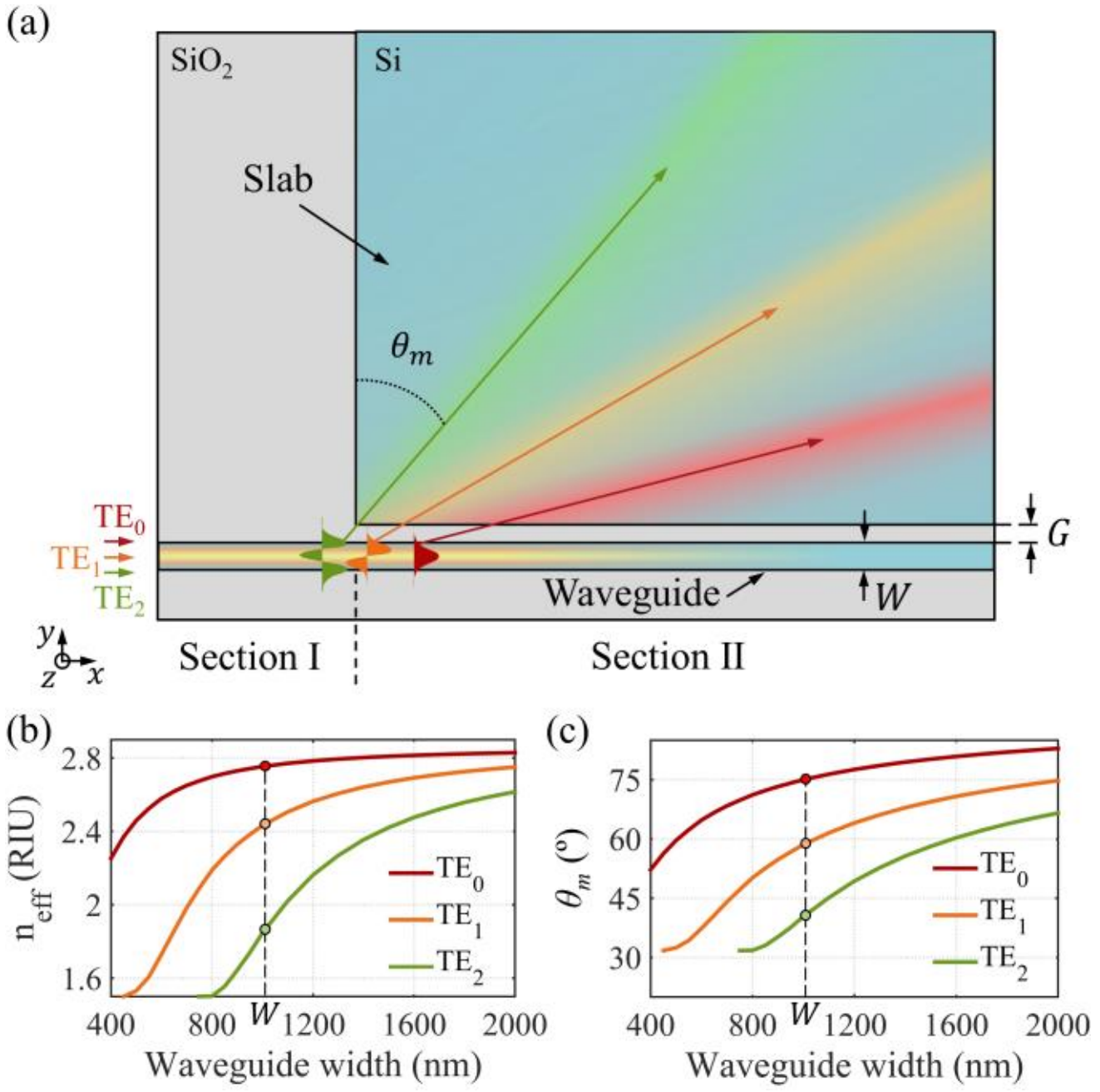

Fig. 1. (a) Schematic top view of a waveguide evanescently coupled to an adjacent slab. A silicon strip waveguide of width $W$ is placed at a distance $G$ from a silicon slab. Fundamental (red), first-order (orange), and second-order (green) TE modes are injected from the left side of the waveguide and propagate along the $x$-axis. Each mode is coupled to a vertically-confined ($z$-axis) beam within the slab with a different propagation angle $\theta_m$. (b) Effective index and (c) propagation angle within the slab of $TE_0$, $TE_1$ and $TE_2$ modes as a function of the waveguide width, calculated for a Si thickness of 220 nm at a wavelength of 1550 nm.

### B. Design of the three-mode demultiplexer

The proposed three-mode demultiplexer is presented in Fig. 2(a). The device comprises an input waveguide of width $W_I$, a coupling region with linearly varying waveguide width and slab separation, and a final section with fixed waveguide width and slab separation of $W_E$ and $G_E$, respectively. Changing the waveguide width in the coupling region results in a gradual variation of the propagation angle of the slab-propagating beam [see Fig. 1(c)], while the change in the slab gap modifies the coupling strength. These two effects are engineered to focus the slab-propagating beams into a near-Gaussian-shaped profile that is coupled to the fundamental mode of a strip waveguide placed at the focal point. This approach allows coupling each mode of the input waveguide to the fundamental mode of a different output waveguide, thereby performing mode demultiplexing and conversion simultaneously. Note that the approach proposed here can be seamlessly extended to handle a larger number of modes.

The device is designed to maximize the coupling efficiency between each mode of the input waveguide and the corresponding output waveguide while maintaining a low crosstalk. Device dimensions are optimized using three-dimensional finite-difference time-domain (3D FDTD) simulations. The input waveguide has a width of $W_I = 1$ µm to support the fundamental, first-order, and second-order modes with transverse-electric (TE) polarization in the wavelength range between 1450 nm and 1650 nm. The dimensions of the coupling region are $G_I = 400$ nm, $G_E = 100$ nm, $W_E = 200$ nm and $L_T = 35$ µm. These dimensions ensure that all the power in the three input waveguide modes is coupled to slab-propagating beams, simultaneously achieving a near-Gaussian profile for the three slab beams. Further details on the design are provided in Supplement 1, Section 1.

As shown in Fig. 2(a), higher-order modes begin to couple to slab beams at early taper positions owing to their weaker modal confinement. The $TE_2$ (green), $TE_1$ (orange) and $TE_0$ (red) modes are completely radiated along the taper as Gaussian-like beams focused on different output points, namely O3, O2 and O1 with respective angles $\theta_2 = 34°$, $\theta_1 = 37°$ and $\theta_0 = 44°$. Figures 2(b), 2(c) and 2(d) show the normalized electric field profile at these focusing points (fixed $y$-position) and along the $x$-direction. The radiated fields are fitted to Gaussians with distinct mode field diameters (MFDs), yielding a high overlap integral of 97% for $TE_0$ input, 88% for $TE_1$ input, and 92% for $TE_2$ input. MFDs used for the Gaussians represented by solid blue curves are 2.4 µm, 3.2 µm and 5.4 µm. The calculated transmittance to each output is shown in Figs. 2(e), 2(f) and 2(g) when $TE_0$, $TE_1$, and $TE_2$ modes are injected into the strip waveguide, respectively. Simulations show insertion losses as low as 0.14 dB at the central wavelength for $TE_0$, with a slight increase to 0.65 dB and 0.69 dB for $TE_1$ and $TE_2$ modes, respectively. The degradation of efficiency with wavelength detuning has its origin in chromatic dispersion, which causes a shift of the focal point. Nevertheless, the demultiplexer exhibits a remarkable low crosstalk within a broad bandwidth of 200 nm. Specifically, the attained crosstalk values are better than -41 dB for $TE_0$, -40 dB for $TE_1$ and -28 dB for $TE_2$ demultiplexing over the entire simulated bandwidth. The effect of potential fabrication imperfections on device performance is discussed in Supplement 1, Section 2.

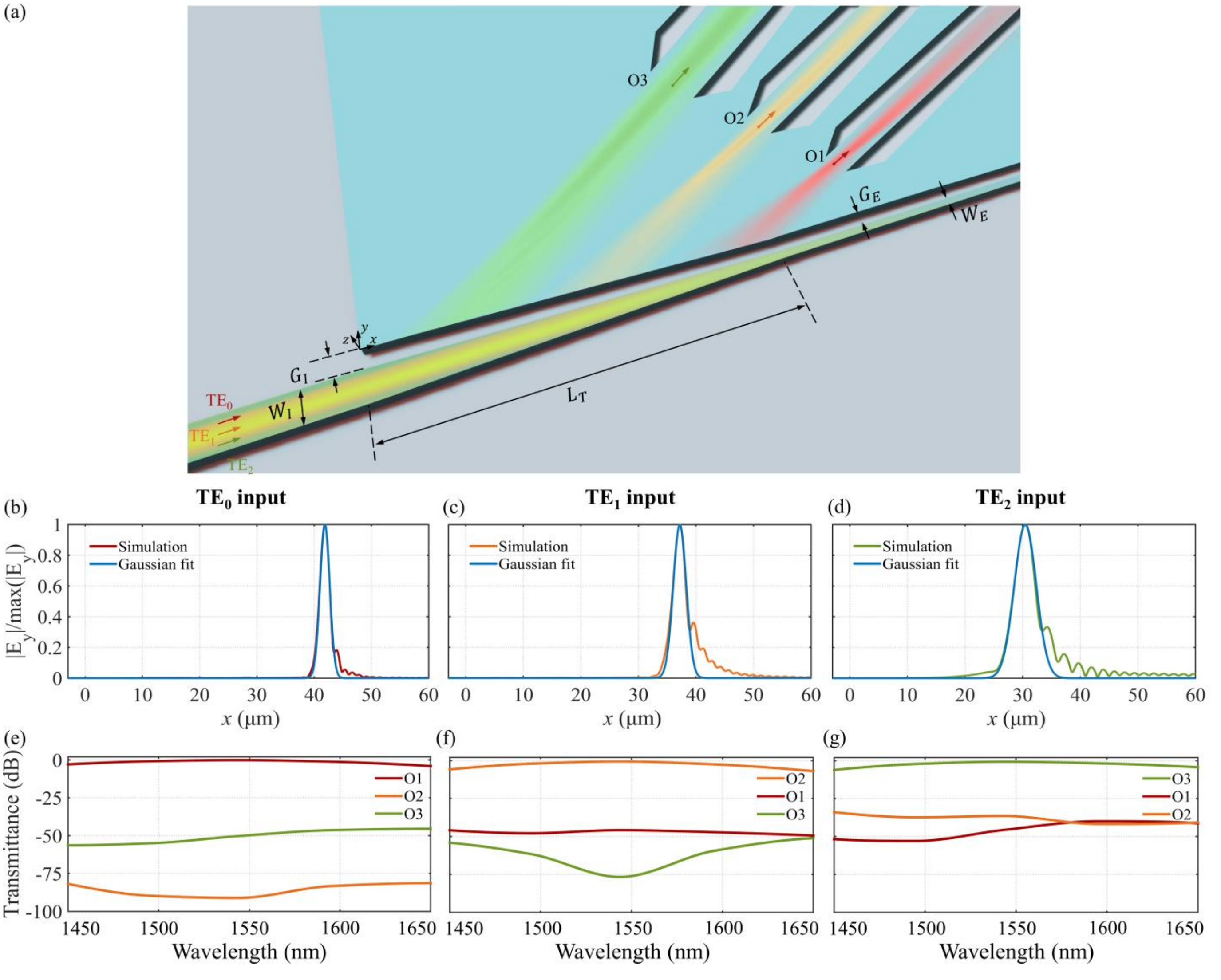

Fig. 2. (a) Three-dimensional schematic of the proposed three-mode demultiplexer comprising a tapered waveguide and an adjacent slab. The PMMA cladding is not shown for clarity. The simulated electric field distribution in the $xy$-plane at mid-height of the Si layer is superimposed on the structure when $TE_0$ (red), $TE_1$ (orange) and $TE_2$ (green) are injected at $\lambda_0 = 1550$ nm. Output waveguide apertures are located at the focal points O$m$+1 to collect the free-propagating beams. Normalized electric field magnitude $|E_y|$ for (b) $TE_0$, (c) $TE_1$ and (d) $TE_2$ input. Simulated results are obtained from the leaked-field distribution within the slab at the optimum position in the $y$-direction for each focal point and along the $x$-axis. The simulated transmittance to each of the output as a function of the wavelength when (e) $TE_0$, (f) $TE_1$ and (g) $TE_2$ are launched into the strip nanowire.

### C. Fabrication and experimental characterization of the MDM link

We implemented an MDM link by connecting two mode (de)multiplexers in a back-to-back configuration. The link comprises three input and three output single-mode waveguides and a central multimode waveguide. The input multiplexer couples the fundamental mode of each input waveguide to a different mode of the multimode section. The output demultiplexer couples each mode of the multimode waveguide to the fundamental mode of one output waveguide. Focusing grating couplers optimized for TE polarization are used to inject and extract the light from the chip with a fiber array. We included a reference waveguide on the outermost part of the test structure. The device was fabricated using a 220-nm-thick single crystal Si layer of an SOI wafer, with a 3-µm-thick buried oxide (BOX) layer. The patterns were defined by electron-beam lithography (RAITH EBPG 5000 Plus) and transferred via reactive ion etching (ICP-DRIE SPTS). Optical and scanning electron microscope (SEM) images were taken prior the deposition of the upper cladding. The sample was then spin coated with a 1.5-µm-thick PMMA. Figure 3(a) shows optical images of the MDM link, with zoomed-in SEM images of the taper-slab coupling region and the collecting output waveguides.

Experimental characterization of the link transmittance is shown in Figs. 3(b), 3(c) and 3(d), when the light is injected through inputs I2, I3, and I4, respectively. The transmittance of each output is obtained by normalizing the measured power at the output ports to the measured power at the reference waveguide output in order to calibrate out the fiber-chip coupling loss. The measured insertion losses are as low as 0.3 dB, 0.9 dB and 1.7 dB at the transmission peak wavelengths with 1-dB bandwidths of 143 nm, 96 nm, and 84 nm for the $TE_0$, $TE_1$, and $TE_2$ mode channels, respectively. Losses of higher-order modes are larger than fundamental mode losses due to slightly lower overlap with the Gaussian-like profile of the collecting output waveguide modes. On the other hand, measured inter-modal crosstalk is below -31.4 dB, -28.3 dB and -25.4 dB at the transmission peak wavelengths for $TE_0$, $TE_1$, and $TE_2$ mode channels, respectively. Additionally, unprecedented crosstalk values reaching -40 dB over a broad bandwidth is observed for $TE_0$ and $TE_2$ channels. Considering 1443 – 1638 nm wavelength range (195 nm bandwidth), the crosstalk is better than -20 dB for all the channels. Transmission peaks of the three channels are slightly shifted towards shorter wavelengths. We attribute this small discrepancy with simulations to intrinsic fabrication variability.

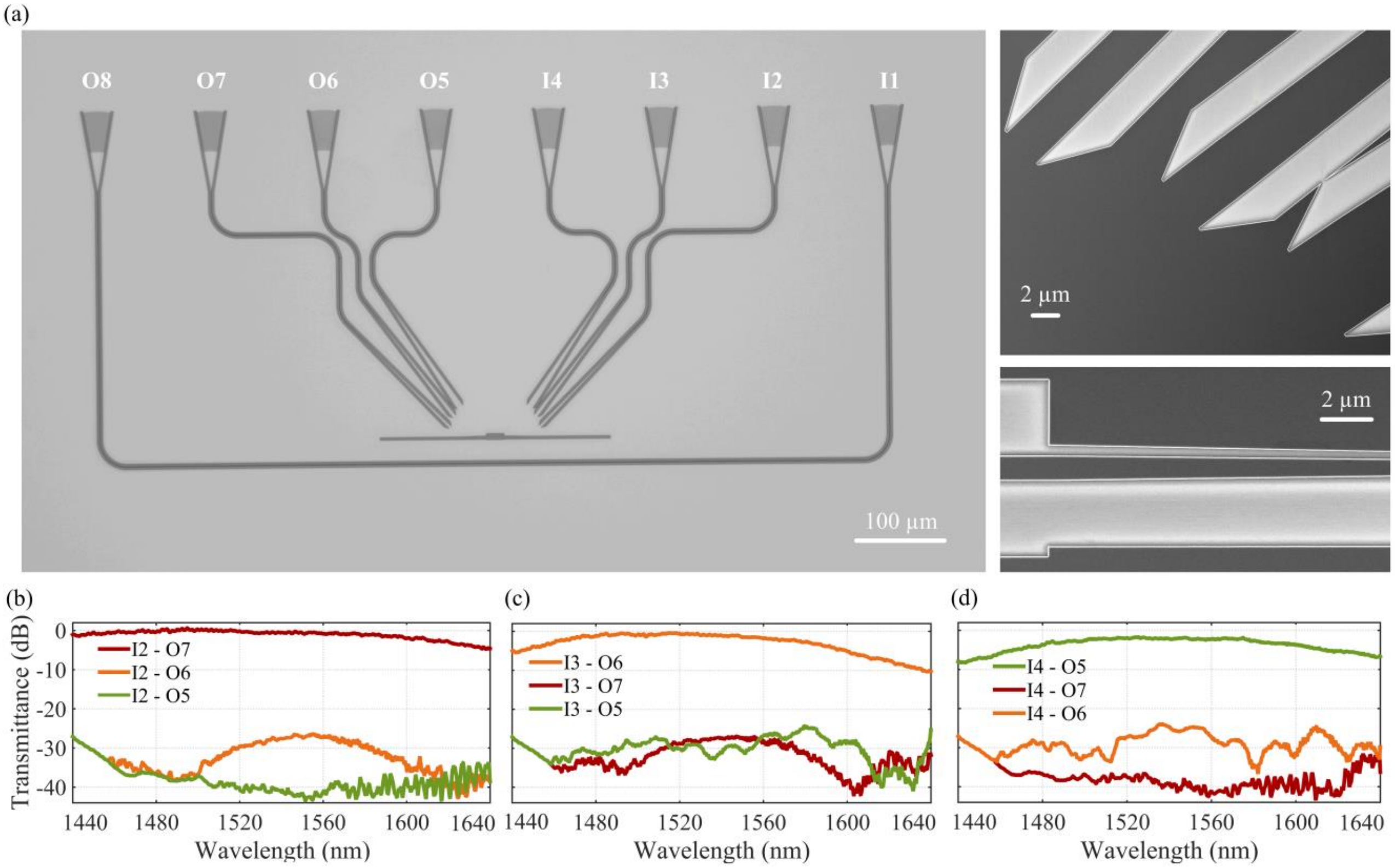

Fig. 3. (a) Optical and scanning electron microscope images of the fabricated MDM link. Input and outputs have been numbered from right to left. The top-right inset shows details of collecting output waveguides, whereas the bottom-right inset shows the taper-slab coupling region. Measured transmission spectra of the complete multiplexer-demultiplexer link for light input at (b) I2, (c) I3, and (d) I4, which correspond to $TE_0$, $TE_1$, and $TE_2$ channels, respectively.

## 4. POLARIZATION DIVISION MULTIPLEXING

### A. Design of the two-polarization demultiplexer

We exploit the proposed approach to demultiplex the two orthogonal TE and transverse-magnetic (TM) polarizations in the waveguide. The proposed two-polarization demultiplexer is schematically shown in Fig. 4(a). In this case, we choose a slot waveguide to achieve sufficient angular separation for the slab beams excited by the fundamental $TE_0$ and $TM_0$ waveguide modes, respectively. The propagation angles for the beams phase-matched to the $TE_0$ and $TM_0$ modes of a strip waveguide, calculated as a function of the waveguide width ($W$), are shown in Fig. 4(b). The angle difference is lower than 10°, hampering the spatial separation of the two beams. The propagation angle of the slab-propagating beam is governed by the ratio between the effective indices of the waveguide and the slab, as dictated by Eq. (2). The effective indices of the $TE_0$ and $TM_0$ modes of a strip waveguide are quite different (e.g., 0.66 for $W = 500$ nm and 1550 nm wavelength). However, the ratio with the slab indices is very similar (e.g., 0.86 for $W = 500$ nm and 1550 nm wavelength), resulting in a comparable propagation angle for the slab-propagating beams. This limitation is overcome using a slot waveguide. We fix a slot width of $G_S = 100$ nm and calculate the propagation angles for the $TE_0$ ($\theta_{TE}$) and $TM_0$ ($\theta_{TM}$) modes of the slot waveguide as a function of the rail width, $W_R$ [see Fig. 4(c)]. For the input waveguide, we choose a rail width of $W_R = 350$ nm, yielding initial propagation angles of $\theta_{TE} = 47°$ and $\theta_{TM} = 60°$.

The rail width and the separation between the slab and the slot waveguide are linearly reduced along the coupling region to achieve a near-Gaussian profile at the focal points for the two slab beams. The optimized geometrical parameters are $G_I = 400$ nm to $G_E = 100$ nm, $W_E = 100$ nm and $L_T = 20$ µm. 3D FDTD simulations are carried out to assess the performance of the polarization beam splitter (PBS). The fields radiated into the slab have an MFD of 4 µm for $TE_0$ input and 6.8 µm for $TM_0$ input. The simulated transfer function for each polarization is shown in Figs. 4(d) and 4(e). Insertion losses are as low as 0.26 dB for the $TE_0$ mode and 0.15 dB for the $TM_0$ mode at the central wavelength. Crosstalk is -61.9 dB and -47.9 dB at the same wavelength when $TE_0$ and $TM_0$ modes are injected, respectively. Notably, the crosstalk is below -37.4 dB for both modes within the simulated bandwidth.

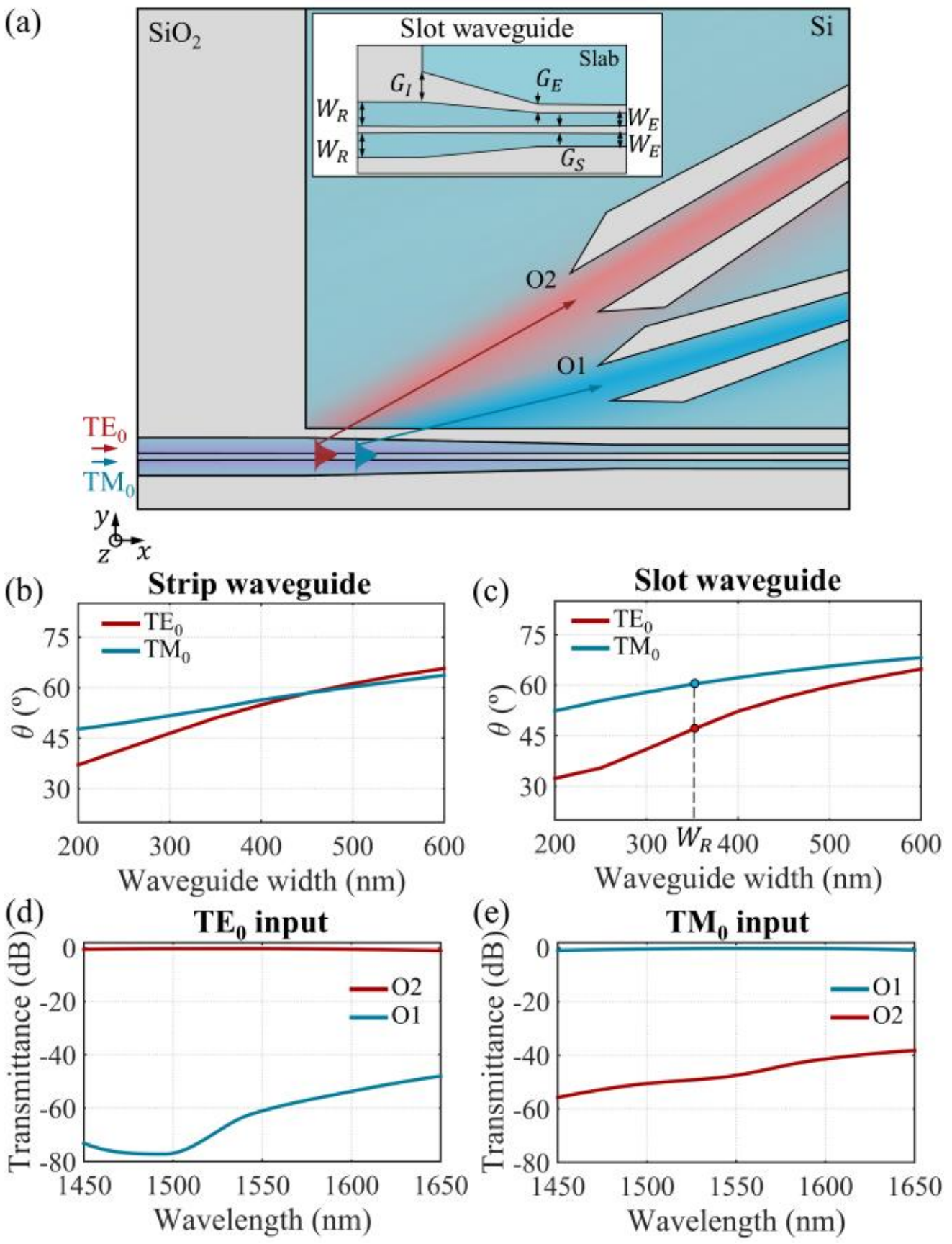

Fig. 4. (a) Schematic top view of the proposed two-polarization demultiplexer comprising a tapered slot waveguide and an adjacent slab. The PMMA cladding is not shown for clarity. The inset shows the geometry of the slot waveguide. Propagation angle within the slab of $TE_0$ and $TM_0$ modes as a function of the waveguide width for (b) a strip waveguide and (c) a slot waveguide, calculated for a Si thickness of 220 nm at $\lambda_0 = 1550$ nm. The simulated transmittance to each of the output as a function of the wavelength when (d) $TE_0$ and (f) $TM_0$ are launched into the slot nanowire.

### B. Fabrication and experimental characterization of the PDM link

We fabricated complete PDM links comprising two polarization beam splitters connected in a back-to-back configuration, using the same SOI wafers and fabrication methods described in Section 3.C. Optical and SEM images of the fabricated devices are shown in Fig. 5(a). Two PDM links with nominally identical dimensions for the multiplexers and different grating couplers were fabricated to characterize losses and crosstalk, respectively. The PDM link used for loss characterization has grating couplers optimized for TE polarization for input 4 and output 5 and grating couplers optimized for TM polarization for input 3 and output 6. The PDM link used for crosstalk characterization has TE grating couplers for input 4 and output 6 and TM grating couplers for input 3 and output 5. TE and TM grating couplers have similar radiation angles. Each link includes two reference waveguides on the outermost part to perform the alignment and transmittance normalization for both TE and TM polarizations.

The transmittance of the link is characterized using the experimental setup described in Supplement 1, Section 3. Measured peak insertion losses are as low as 0.5 dB with a 1-dB bandwidth exceeding 100 nm and crosstalk of -40.1 dB for $TE_0$, and as low as 0.7 dB with a 1-dB bandwidth exceeding 108 nm and a crosstalk of -46.7 dB for $TM_0$. Ultra-low inter-modal crosstalk of <-35 dB is attained for both polarizations within the measured 180-nm bandwidth. Crosstalk reaching values below -40dB could be observed in the 1542 – 1680 nm wavelength range for the $TM_0$ channel.

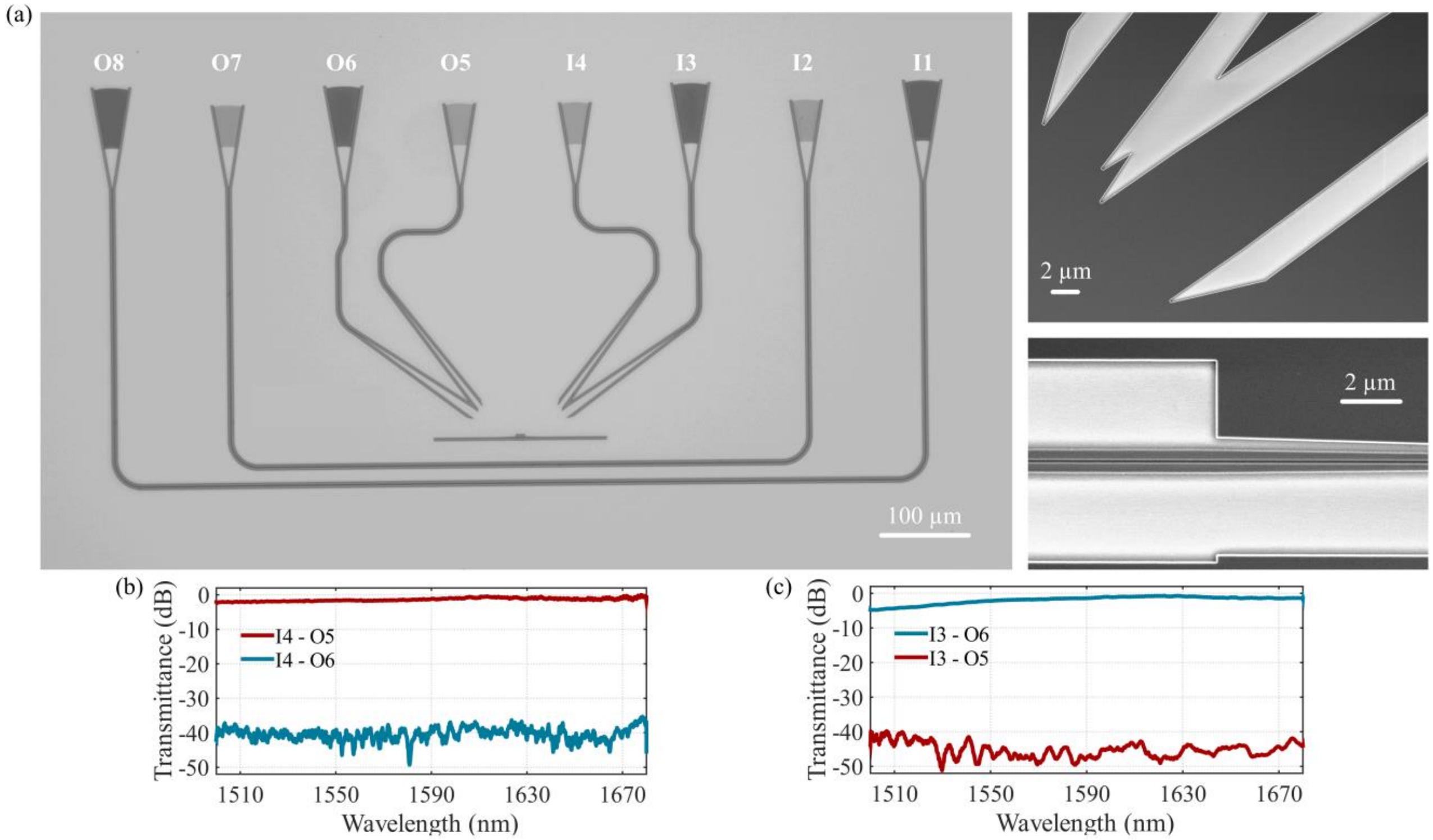

Fig. 5. (a) Optical and scanning electron microscope images of one of the fabricated PDM links. Input and outputs have been numbered from right to left. The top-right inset shows details of collecting output waveguides, whereas the bottom-right inset shows the slot-slab coupling region. Measured transmission spectra of the complete multiplexer-demultiplexer link for light input at (b) I4 and (c) I3, which correspond to $TE_0$ and $TM_0$ channels, respectively.

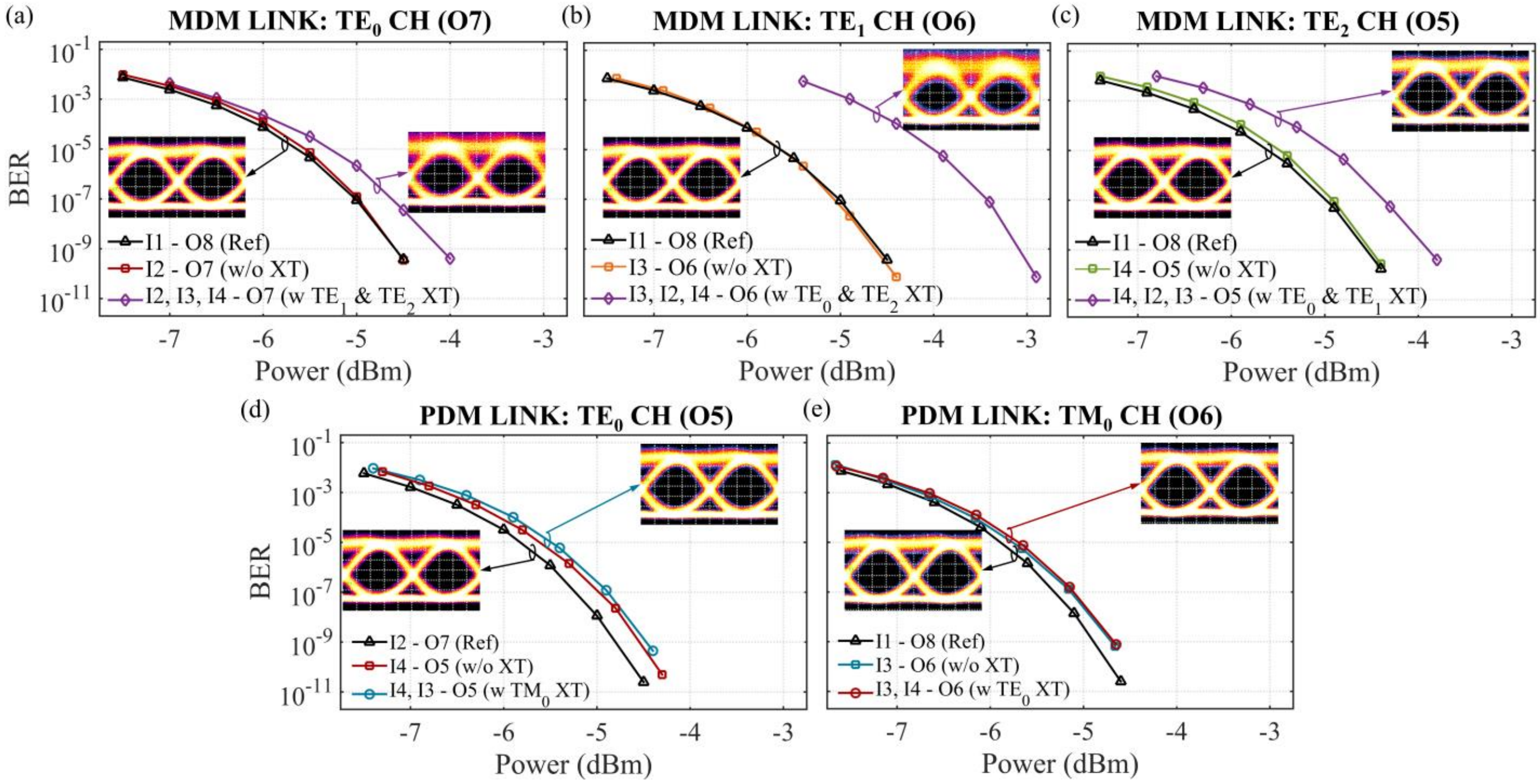

Fig. 6. Crosstalk penalty assessment of MDM and PDM links for a transmission bit rate of 40 Gbps at 1549 nm. Bit-error-rate measurements as a function of the received power for (a) $TE_0$ channel, (b) $TE_1$ channel and (c) $TE_2$ channel of the MDM link, and for (f) $TE_0$ channel and (e) $TM_0$ channel of the PDM link. The insets show the corresponding eye diagrams of the demultiplexed signals ($x$-axis: 5ps/div and $y$-axis: 0.5 mV/div). Ref, Reference; XT, Crosstalk; CH, Channel.

## 5. HIGH-SPEED DATA TRANSMISSION

We have characterized the data transmission performance of the three-mode and two-polarization multiplexed links in terms of bit-error-rate. The experimental setup employed is described in Supplement 1, Section 3. The figure of merit used to quantify the dynamic performance is the power penalty, defined as the difference in optical power measured in the receiver with and without signal impairment (i.e., crosstalk), for error-free transmission (BER<$10^{-9}$) without using any correction technique. Figure 6 shows the evolution of the BER as a function of the received power for the reference, single-port transmission, and MDM and PDM operation. For the MDM link, the worst case is when all three signals are introduced simultaneously due to the crosstalk of the aggressor channels. Still, the power penalties are as low as 0.5 dB, 1.5 dB and 0.6 dB for $TE_0$, $TE_1$ and $TE_2$ mode channels, respectively. As expected from static measurements, the highest power penalty is obtained for $TE_1$ mode channel since the crosstalk of the aggressor channels towards this other channel is higher [see Fig. 3(b) and 3(d), orange curves]. The PDM link, on the other hand exhibits negligible power penalties for both $TE_0$ (0.08 dB) and $TM_0$ (0.03 dB) channels. Power difference between reference (Ref) and single-port (w/o XT) measurements are 0 dB ($TE_0$ channel), 0.04 dB ($TE_1$ channel) and 0.06 dB ($TE_2$ channel) for the MDM link, and 0.25 dB ($TE_0$ channel) and 0.19 dB ($TM_0$ channel) for the PDM link. In all cases, the demultiplexed signals exhibit clear and open eye diagrams indicating a low effective crosstalk, as shown in the insets of Fig. 6. These results indicate that the proposed devices have an excellent potential for error-free data transmission, paving the way for next-generation MDM and PDM communication applications.

## 6. DISCUSSION AND CONCLUSIONS

In conclusion, we have shown a new approach for spatial and polarization multiplexing, exploiting the excitation of optical beams free-propagating in the chip plane to achieve state-of-the-art performance. To demonstrate the concept, we have developed a three-mode and a two-polarization links allowing error-free propagation of 40 Gbps signals with negligible power penalties. The three-mode link takes advantage of the strong modal dispersion in strip waveguides to realize mode multiplexing and conversion based on free-space-like optical beam forming on chip. The proposed three-mode link comprising a multiplexer and a demultiplexer shows a measured crosstalk lower than -20 dB over a 195 nm bandwidth (1443 – 1638 nm) that fully covers the S, C and L telecommunication bands, and partially covers the E and U bands. Furthermore, insertion losses lower than 1.7 dB with a 1-dB bandwidth of 84 nm are attained for all three modes. The two-polarization link harnesses birefringence engineering in slot waveguides to yield an ultra-low crosstalk below -35 dB within the 1500 – 1680 nm wavelength range for both polarizations (covering the entire C, L, and U bands, and part of the S band). Measurements also showed low insertion losses (<0.7 dB) for $TE_0$ and $TM_0$, with a 1-dB bandwidth exceeding 100 nm, limited at the upper bound by the wavelength range of the laser available in our setup.

The low crosstalk values (<-40 dB) observed in the MDM link for $TE_0$ and $TE_2$ modes suggest that increasing the separation of the focal points within the slab is a simple but effective way to reduce crosstalk. This could be achieved by increasing the taper length to increase the focal length or by bending the coupling region. On the other hand, insertion losses of higher-order modes could be further reduced by improving the overlap between the modes coupled to the slab free-propagation region and the Gaussian-like profile of the collecting output waveguide modes, for example by implementing a nonlinear taper in the coupling region. Nevertheless, the proposed devices are, to the best of our knowledge, among the mode multiplexers and polarization beam splitters with lowest measured crosstalk within an ultra-broad bandwidth. For the sake of comparison, Table 1 shows the performance of demonstrated state-of-the-art three-channel MDM links and polarization beam splitters. Note that state-of-the-art polarization beam splitters have been usually measured in standalone configuration, therefore, we have halved the insertion losses measured by our PDM link to ensure a fair comparison.

**Table 1. Comparison of Demonstrated State-of-the-art Three-channel MDM Links and PBS[a]**

| MDM/PBS | Ref. | Architecture | IL [dB] | $BW_{CT<-20\,dB}$ [nm] | $BW_{CT<-30\,dB}$ [nm] | L [μm] |
|---|---|---|---|---|---|---|
| MDM | [7] | Asymmetric Y-junction | 3.3 ~ 5.7 | - | - | 320 |
| | [9] | Counter-tapered coupler | ~10* | - | - | 200 |
| | [10] | Adiabatic coupler | ~0.2 | 65 | - | 310 |
| | [11] | Subwavelength asymmetric Y-junction | <2.5 | ~50* | - | 4.8 |
| | [12] | Subwavelength DC | <3.8 | - | - | >100* |
| | This work | In-slab beamforming | 0.3 ~ 1.7 | 195 | - | 35 |
| PBS | [14] | MMI | 1.2 ~ 2.2- | - | - | 132.64 |
| | [15] | MZI | ~15 | ~20* | - | 340 |
| | [16] | DC | <0.5 | 125 | - | 97.4 |
| | [17] | ADC | - | - | - | 25.5 |
| | [18] | Counter-tapered coupler | <0.5 | 15* | - | 5 |
| | [19] | Bent DC | <0.35 | 135 | 70 | 20 |
| | [20] | Cascaded triple bent DC | <0.8 | 90 | ~50 | ~26 |
| | [21] | MMI with photonic crystal | <0.81 | 77 | - | 71.5 |
| | [22] | Slot ADC | - | - | - | >13.6 |
| | [23] | Subwavelength MMI | <2.5 | - | - | 92.7 |
| | [24] | Tilted-nanograting | <1.5* | - | - | >6.8 |
| | This work | In-slab beamforming | 0.3 ~ 0.4 | >180 | >180 | 20 |

[a]Values marked with an asterisk correspond to values estimated from figures. IL, Insertion Loss; CT, Crosstalk; BW, Bandwidth; L, Length.

A high-speed optical communications demonstration was also performed to illustrate the applicability of the proposed devices. System-level experiments at 40 Gbps without forward error correction were conducted for both MDM and PDM links, showing clear and open eye diagrams during the joint transmission of data channels. BER measurements further validated the good transmission capabilities with less than a 1.5 dB power penalty for the MDM link and below 0.08 dB for the PDM link. The ease of scalability of the proposed architecture along with such low penalties can be exploited to implement hybrid WDM-PDM-MDM optical links to significantly enhance the transmission capacity.

The strategy demonstrated here could seamlessly be extended for a larger number of modes and simultaneous modal and polarization multiplexing. The low crosstalk, wide bandwidth and versatility of the proposed multiplexing approach presented in this work will open up unique possibilities in quantum information sciences, optical sensing, on-chip wireless communications and nonlinear photonics.

**Funding.** French Industry Ministry (Nano2022 project under IPCEI program); Agence Nationale de la Recherche (ANR-MIRSPEC-17-CE09-0041); European Union's Horizon Europe (Marie Sklodowska-Curie grant agreement Nº 101062518) The fabrication of the device was performed at the Plateforme de Micro-Nano-Technologie/C2N, which is partially funded by the Conseil General de l'Essonne. This work was partly supported by the French RENATECH network.

**Disclosures.** The authors declare no conflict of interest.

**Data availability.** Data underlaying the results presented in this paper are not publicly available at this time but may be obtained from the authors upon reasonable request.

**Supplemental document**. See Supplement 1 for supporting content.

# Spatial and polarization division multiplexing harnessing on-chip optical beam forming: supplemental document

## 1. Design of the coupling region for the mode and polarization demultiplexers

The coupling region is designed using two-dimensional finite-difference time-domain (2D FDTD) simulations combined with the effective index method. Initial strip and slot waveguide dimensions are chosen to support three transverse-electric (TE) modes and one TE and one transverse-magnetic (TM) modes, respectively. The initial gap ($G_I$) is chosen to minimize back-reflections at the boundary between the isolated waveguide and the waveguide coupled to the adjacent slab, resulting in $G_I = 400$ nm for the three-mode demultiplexer and $G_I = 400$ nm for the two-polarization demultiplexer. Final widths and separations are chosen to keep a minimum feature size of 100 nm. We sweep the taper length to ensure high transmission for demultiplexing of each mode, and more importantly, to verify the distance between the focal points of beams in the slab (also referred to as focal separation). The latter is crucial to place the collecting output waveguides on the slab without overlapping structures, power coupling or high crosstalk.

Figure S1(a) shows the transmittance as a function of the taper length for the design of the three-mode demultiplexer. Insertion losses of less than 1 dB can be achieved for all three modes with a taper length of 16.4 µm. However, the separation at adjacent focal points (i.e., between $TE_0$ and $TE_1$, and between $TE_1$ and $TE_2$) is only 5.3 µm. To ensure that there will be no overlap between the collecting waveguides, and to decrease crosstalk and insertion losses, we choose a taper length of 30 µm. For this value, the insertion losses are 0.06 dB for $TE_0$, 0.2 dB for $TE_1$, and 0.64 dB for $TE_2$, while the focal spacing is increased by 5 µm (to 10.3 µm).

A similar strategy is followed for the design of the polarization demultiplexer. In this case, maximum transmission is achieved for a taper length of 15 µm. However, due to the larger mode field diameter of the radiated fields (e.g., 3.4 µm for $TE_0$ input and 6.4 µm for $TM_0$) and the small variation of transmission with taper length, a tapered slot length of 20 µm is chosen increasing the focal separation from 10.6 µm to 13.2 µm. For the selected value, the $TE_0$ and $TM_0$ mode insertion losses drop slightly to 0.07 dB and 0.21 dB, respectively.

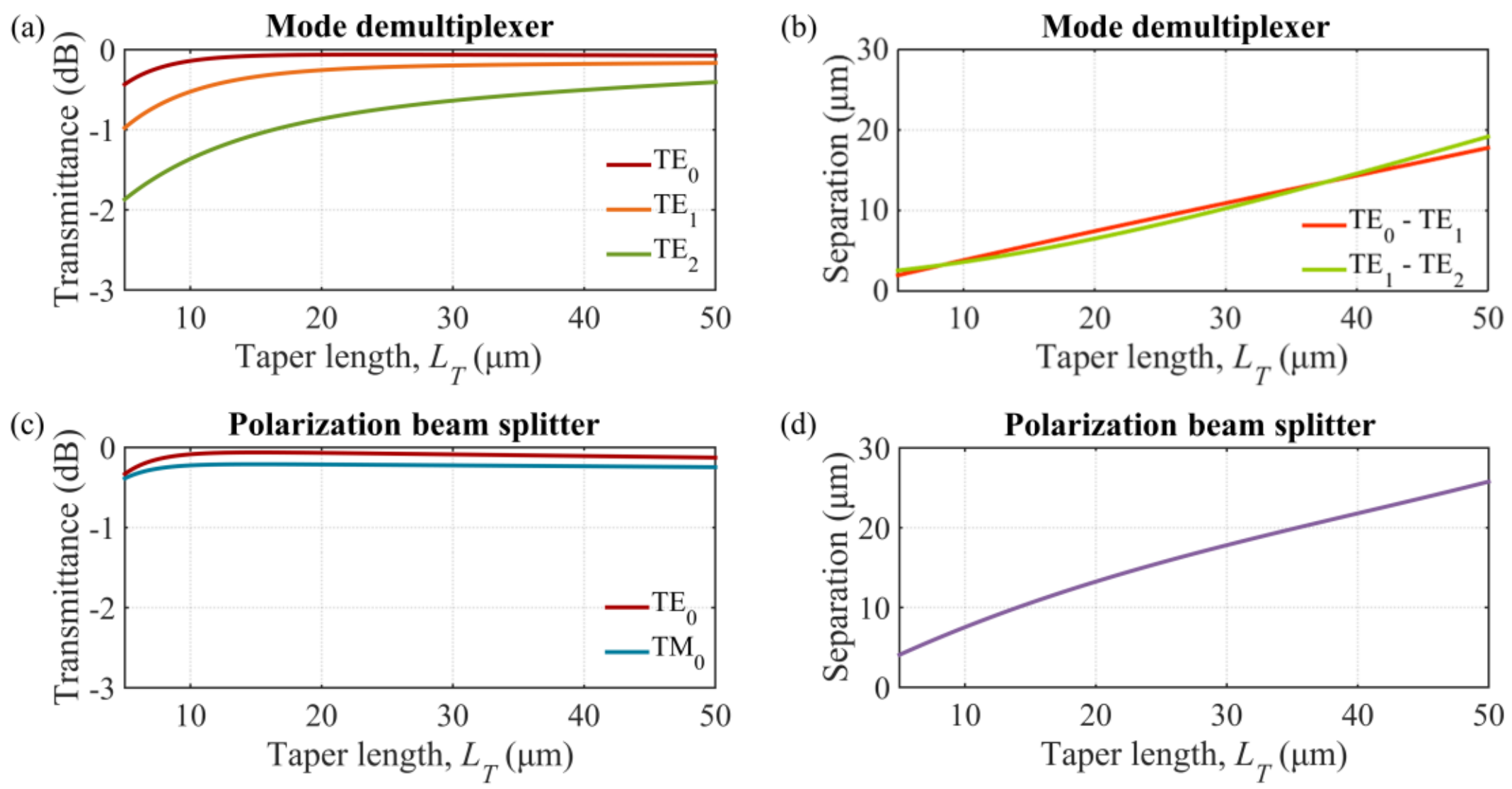

Fig. S1. Simulated (a) transmittance and (b) focal separation as a function of the taper length $L_T$ for the three-mode demultiplexer. Simulated (c) transmittance and (d) focal separation as a function of the taper length $L_T$ for the polarization beam splitter. Calculations are performed at the wavelength of $\lambda_0 = 1550$ nm.

## 2. Study of fabrication tolerances

Tolerances to fabrication deviations are calculated using three-dimensional FDTD method for both the three-mode demultiplexer (see Fig. S2) and the polarization beam splitter (see Fig. S3). To mimic typical fabrication imperfections produced during the etching process, we modify the width of the waveguide and slab by also affecting the separation between the tapered waveguide and the slab (i.e., the central position of the waveguide and slab are maintained, and their width are widened or narrowed by $\Delta W$). The thickness of the silicon layer can also vary across the wafer, so we also study how a modified silicon height (by $\Delta H$) affects device performance.

From the simulated spectra for the three-mode demultiplexer shown in Fig. S2, a crosstalk lower than -25.7 dB can still be obtained over the simulated 200 nm bandwidth, which corresponds to the case of $\Delta W = +10$ nm for $TE_2$ input [see Fig. S2(c)]. The main effect of positive deviations in silicon width and height is a slight shift of the transmission peak towards longer wavelengths, while negative deviations produce the opposite effect. The largest shift of the transmission peak occurs for the $TE_0$ mode with 40 nm and 55 nm for width errors of $\Delta W = +10$ nm and $\Delta W = -10$ nm, respectively. Nevertheless, the insertion losses at these peak wavelengths hardly suffer any degradation, reaching values of 0.2 dB for $TE_0$ ($\Delta W = -10$ nm), 0.71 dB for $TE_1$ ($\Delta W = -10$ nm), and 0.81 dB for $TE_2$ ($\Delta H = -10$ nm) considering for each mode the worst case among the deviations analyzed.

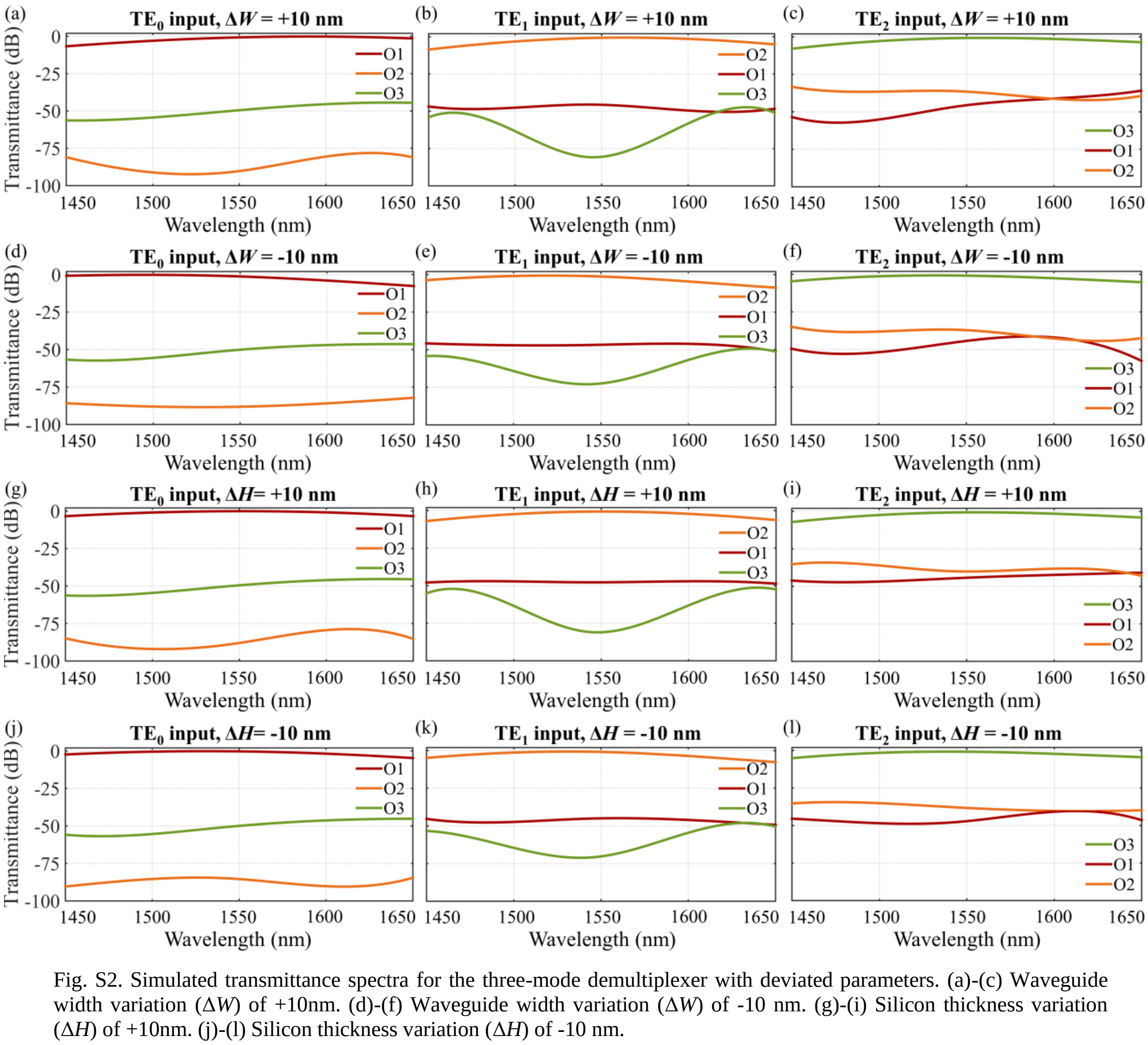

Fig. S2. Simulated transmittance spectra for the three-mode demultiplexer with deviated parameters. (a)-(c) Waveguide width variation ($\Delta W$) of +10nm. (d)-(f) Waveguide width variation ($\Delta W$) of -10 nm. (g)-(i) Silicon thickness variation ($\Delta H$) of +10nm. (j)-(l) Silicon thickness variation ($\Delta H$) of -10 nm.

The study of the polarization beam splitter fabrication tolerances shows that the crosstalk between $TE_0$ and $TM_0$ modes is lower than -35.2 dB in all cases over the entire simulated bandwidth [see Fig. S3(f), corresponding to $\Delta H = +10$ nm for $TM_0$ input]. Similar to the mode demultiplexer, the different deviations in the transverse dimensions of the waveguide and slab result in a transmission peak shift at which maximum insertion losses are 0.71 dB for $TE_0$ input ($\Delta W = -10$ nm) and 0.9 for $TM_0$ input ($\Delta H = -10$ nm) mode. The largest transmission peak shift occurs for $TE_0$ mode (100 nm) for width errors of $\Delta W = \pm 10$ nm.

Results shown in Figs. S2 and S3 show that low crosstalk can still be achieved for the proposed devices even with width and thickness variations as large as ±10 nm.

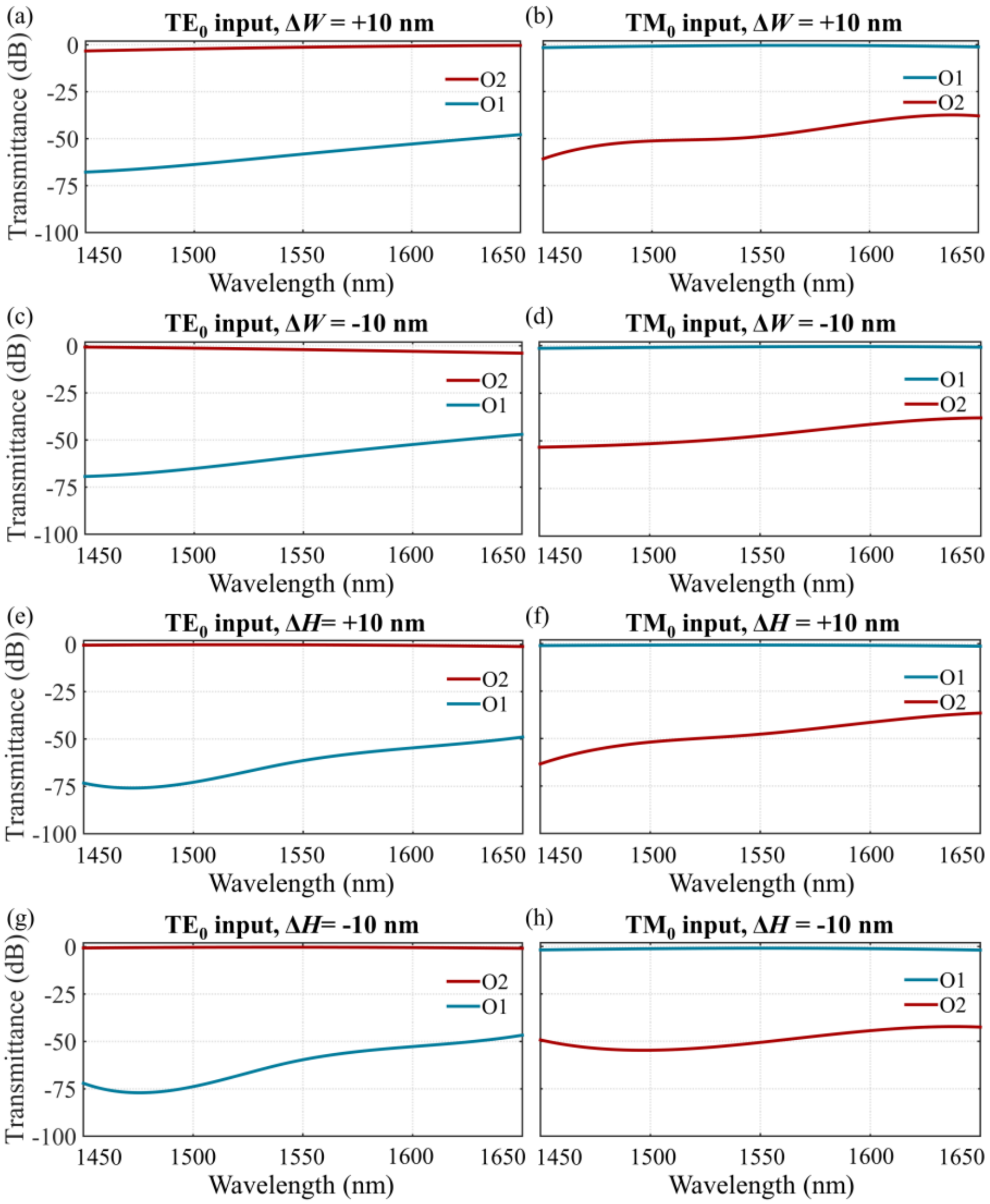

Fig. S3. Simulated transmittance spectra for the polarization beam splitter with deviated parameters. (a)-(c) Waveguide width variation (Δ*W*) of +10nm. (d)-(f) Waveguide width variation (Δ*W*) of -10 nm. (g)-(i) Silicon thickness variation (Δ*H*) of +10nm. (j)-(l) Silicon thickness variation (Δ*H*) of -10 nm.

## 3. Characterization setups

Figure S4(a) shows the schematic of the setup used for the static characterization of the mode-division multiplexing and polarization-division multiplexing links.

For the mode-division multiplexing (MDM) link, we use two continuous-wave tunable external-cavity lasers (Yenista Tunics) connected to a data acquisition system (Yenista CT400) to sweep the 1440 – 1640 nm wavelength range. A three-paddle polarization controller is employed to inject TE-polarized light into the chip and the tilt angle of the fiber array is adjusted sequentially to perform the measurements in different spectral regions.

In order to characterize the transmission response for the polarization-division multiplexing (PDM) link, we employed the same experimental setup as for the MDM link but replaced the Yenista Tunics tunable laser by a high-power laser (Yenista T100S-HP) covering the 1500 – 1680 nm wavelength range to increase dynamic range and to ensure that the measurements are not limited by the noise floor.

On the other hand, the evaluation of bit-error-rates (BER) and eye diagrams is performed using the setup illustrated in Fig. S4(b). Light from a distributed feedback laser emitting at a wavelength of 1549 nm is modulated with pseudo-random binary-sequence (PRBS) pattern on-off keyed data in a non-return-to-zero (NRZ) format by an external modulator. The length of the PRBS patterns is $2^7$-1 and the data rate of the pattern generator is set to 40 Gbps, which corresponds to its upper limit. Then, light is amplified using an erbium-doped fiber amplifier (EDFA) and band-pass filtered to reduce the spontaneous emission noise. Two 50:50 coupler stages are cascaded (C1, C2 and C3) to generate four identical signals of approximately equal power. In this way, three signals could be connected to the device under test (DUT) and the fourth is used as reference to display the modulated input signal on one channel of a high-speed sampling oscilloscope (CSA spectrum analyzer, 60 GHz photodiode). To ensure the correct characterization of the DUT, the decorrelation of the data channels coupled to the device is performed by including a 1-km and 2-km-long optical fibers. A 1 m jumper is used on the remaining input to compensate for the connector losses. The polarization of each data channel is controlled with a polarization controller (PC) to inject TE or TM polarized light into the chip. The signal at the chip output is recovered one at a time and passes through a first 90:10 coupler (C4), where 10% of the power is directed to an optical spectrum analyzer (OSA) for verification purposes. The remaining 90% is amplified and filtered again, and split by a second 90:10 coupler (C5). In this case, 10% of the power distributed by the coupler C5 is amplified and sent to the other channel of the sampling oscilloscope (CSA spectrum analyzer, 40 GHz photodiode) that gives the reconstructed eye diagram. The 90% power after C5 passes through a variable optical attenuator (VOA) and another 90:10 coupler (C6), whose 90% output is routed to the receiver (RX) for BER evaluation and the 10% output is used to control the received power level with a power meter. Detection at the receiver is performed without the use of a limiting amplifier, optical preamplifier, or automatic gain control so that the assessment of transmission penalty is not minimized. To perform the measurements under the same conditions, the main signal always comes from the input connected to the 1 km optical fiber. This means that for the MDM link, the signal from the 1-km-long fiber is connected to input I2 for the characterization of the $TE_0$ channel, to input I3 for the $TE_1$ channel and to input I4 for the $TE_2$ channel. The same procedure is used for the characterization of the PDM link.

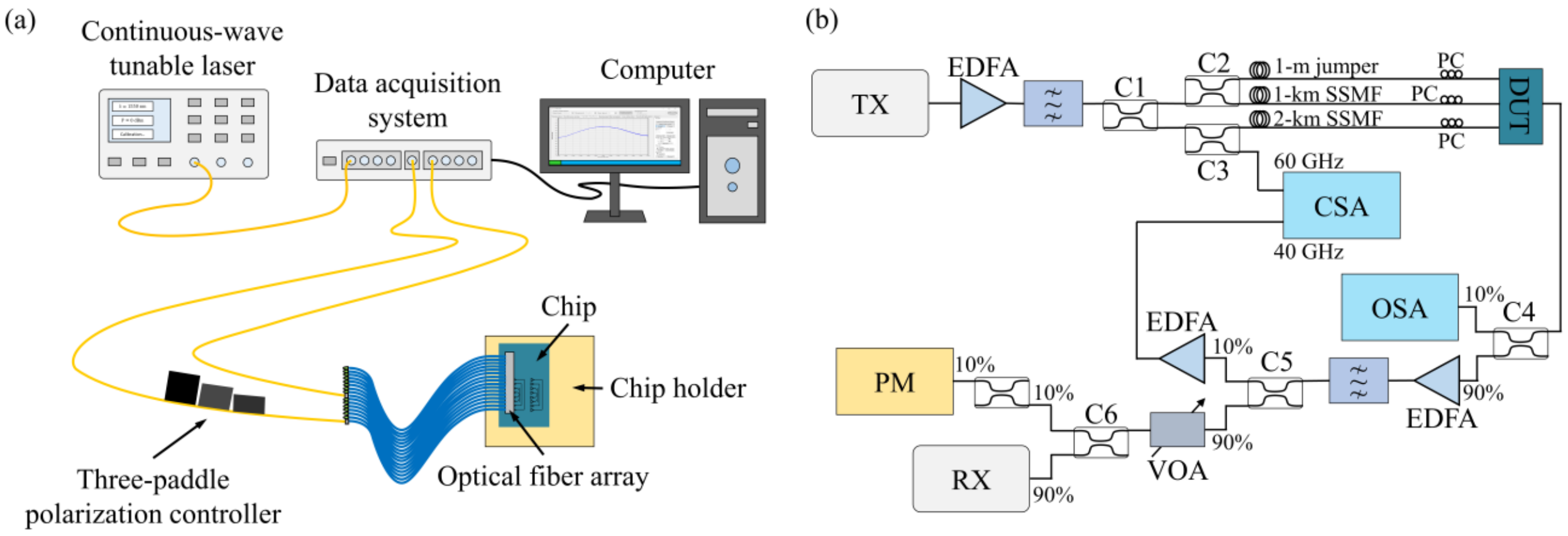

Fig. S4. Schematics of the setups used to perform (a) static and (b) dynamic characterization of MDM and PDM links. TX, transmitter; EDFA, erbium-doped fiber amplifier; CX, coupler X, SSMF, standard single-mode fiber; PC, polarization controller; DUT, device under test; CSA, communications signal analyzer; OSA, optical spectrum analyzer; VOA, variable optical attenuator; PM, power meter; RX, receiver.